\documentclass[twocolumn,times,twocolappendix]{aastex701}

\usepackage{amsmath, latexsym, color, verbatim}

\shorttitle{Stellar winds of O-type stars with JWST/MIRI}
\shortauthors{Hawcroft et al. 2026}

\newcommand{\Mdot}{\hbox{$\dot M$}}

\newcommand{\vinf}{$v_{\infty}$}

\newcommand{\ncrit}{\hbox{$n_{\hbox{\scriptsize \rm crit}}$}}

\newcommand{\kms}{\textrm{km~s}\ensuremath{^{-1}\,}}

\newcommand{\nevi}{\textrm{[Ne\,{\sc vi}]}}
\newcommand{\nev}{\textrm{[Ne\,{\sc v}]}}

\newcommand{\ofour}{\textrm{[O\,{\sc iv}]}}

\def\one{\,{\sc i}}             
\def\two{\,{\sc ii}}
\def\three{\,{\sc iii}}
\def\four{\,{\sc iv}}
\def\five{\,{\sc v}}
\def\six{\,\textsc{vi}}

\usepackage{lineno}
\linenumbers

\begin{document}

\title{Stellar winds of O-type stars traced by high ionization fine-structure emission lines with JWST/MIRI}


\author[0000-0003-0145-8964]{Calum Hawcroft}
\affiliation{Space Telescope Science Institute, 3700 San Martin Drive, Baltimore, MD, 21218, USA}
\email[show]{chawcroft@stsci.edu}

\author[0000-0002-9402-186X]{David R.\ Law}
\affiliation{Space Telescope Science Institute, 3700 San Martin Drive, Baltimore, MD, 21218, USA}
\email{dlaw@stsci.edu}

\author[0000-0002-0806-168X]{Linda J. Smith}
\affiliation{Space Telescope Science Institute, 3700 San Martin Drive, Baltimore, MD, 21218, USA}
\email{lsmith@stsci.edu}

\author[0000-0003-2429-7964]{Alexander W. Fullerton}
\affiliation{Space Telescope Science Institute, 3700 San Martin Drive, Baltimore, MD, 21218, USA}
\email{fullerton@stsci.edu}


\author[0000-0001-5340-6774]{Karl D.\ Gordon}
\affiliation{Space Telescope Science Institute, 3700 San Martin Drive, Baltimore, MD, 21218, USA}
\email{kgordon@stsci.edu}




\author[0000-0001-6000-6920]{Paul A.\ Crowther}
\affiliation{Astrophysics Research Cluster, School of Mathematical \& Physical Sciences, Hounsfield Road, University of Sheffield, Sheffield, S3 7RH, United Kingdom}
\email{paul.crowther@sheffield.ac.uk}

\author[0000-0001-9462-5543]{Marjorie Decleir*}
\affiliation{European Space Agency (ESA), ESA Office, Space Telescope Science Institute, 3700 San Martin Drive, Baltimore, MD 21218, USA}
\altaffiliation{ESA Research Fellow}
\email{mdecleir@stsci.edu}



\author[0000-0002-8163-8852]{Sascha T.\ Zeegers}
\affiliation{SRON Netherlands Institute for Space Research, Niels Bohrweg 4, 2333 CA Leiden, The Netherlands}
\affiliation{Anton Pannekoek Institute for Astronomy, Universiteit van Amsterdam, Science Park 904, 1098 XH Amsterdam, The Netherlands}
\email{zeegers at strw.leidenuniv.nl}

\author[0000-0003-1299-8878]{Christiana Erba}
\affiliation{Space Telescope Science Institute, 3700 San Martin Drive, Baltimore, MD, 21218, USA}
\email{cerba@stsci.edu}

\author[0000-0002-7204-5502]{Richard Ignace}
\affiliation{Department of Physics \& Astronomy, East Tennessee State University, Johnson City, TN37614, USA}
\email{ignace@mail.etsu.edu}

\author[0000-0001-5094-8017]{D. John Hillier}
\affiliation{Department of Physics and Astronomy \& Pittsburgh Particle Physics, Astrophysics, and Cosmology Center (PITT PACC), University of Pittsburgh, 3941 O’Hara Street, Pittsburgh, PA 15260, USA}
\email{hillier@pitt.edu}

\begin{abstract}

We investigate the presence of high ionization fine-structure emission lines across a range of 22 OB-type stars observed with JWST-MIRI as part of calibration programmes and the WISCI and MEAD projects. MIR wind emission is detected in 4 late O-dwarfs (O8~V -- O9~V), 1 early O-dwarf (O5~V) and there are tentative detections in an additional 3 stars (O8.5~II, O8.5~IV and O9~I). We measure the wind speeds and make estimates of lower limits on the mass-loss rates of 5 O-type stars from broad, flat-topped emission in the fine-structure line of \nev\ 14.3\micron. We find terminal wind speeds that are generally in agreement with empirical trends, but note that in some cases the wind speeds are surprisingly low. 
We highlight two main takeaways from this sample, which combine to establish an exciting new window into the winds of massive stars. First, a new diagnostic ability is gained from lines formed at much higher ionization, larger spatial extent, and longer wavelengths than typical wind diagnostics. Secondly, there is frequent incidence of MIR emission in O-type stars, even in the `weak-wind' regime where wind emission is often not detected in the UV and optical. 

\end{abstract}



\section{Introduction}

The radiation-driven winds from hot, massive stars are an effective mechanism for removing material from the stellar surface and ejecting it into the ISM. The mass loss from stellar winds shapes the evolutionary pathways of massive stars, influencing their ultimate fates as supernovae and compact objects \citep{Langer2012}. The magnitude of the winds directly determines the impact of stellar feedback, which is essential for regulating galaxy evolution and interpreting observational signatures \citep{Martins2010, Kennicutt2012, Steidel1996, Leitherer2020}. Consequently, accurate characterization of stellar winds is essential for modeling a wide range of astrophysical processes. Two crucial parameters used to quantify stellar winds are the terminal velocity and mass-loss rate, both of which are sensitive to fundamental stellar parameters such as mass, luminosity \added{and metallicity}. 

For O-type stars, the wind driving scenario through radiation pressure on metal ion transitions is well established, but complexities associated with wind structure and opacity introduce uncertainties in our ability to use spectral diagnostics to interpret stellar wind signatures \citep{Kudritzki2000, Puls2008, Vink2022}. For example, over-dense clumps of wind material arising due to the inherent line-deshadowing instability result in enhanced emission from density-squared dependent recombination processes \citep{Eversberg1998, Fullerton2006}. This limits our ability to constrain mass-loss rates from optical hydrogen and helium lines alone. These lines are also formed close to the stellar photosphere while the wind is still accelerating, so they do not probe the maximum wind velocity. UV metal resonance lines, which scale linearly with density and form further out in the wind, are a more reliable source of wind information for massive stars \citep{Hillier2020}. However, there are also issues with typical UV diagnostics, related to the overall wind opacity \citep{Oskinova2007, Surlan2013, Hawcroft2021, Hawcroft2024}. Additionally, they are no longer sensitive to mass loss below a characteristic luminosity (log$(L/L_{\odot} ) \sim 5.2$) at the onset of the 'weak-wind' regime \citep{Bouret2003, Martins2004, Martins2005b, Marcolino2009, deAlmeida2019}.

A prospective new spectral window has appeared in the mid-infrared (MIR). Emission lines from fine structure transitions of \ofour\ 25.9\micron, \nev\ 14.3\micron\ and \nevi\ 7.6\micron\ have been detected in MIR observations of the O9~V star 10~Lac, which provided direct wind constraints in the weak-wind regime where typical UV and optical diagnostics are limited \citep{Law2024}. These emission lines are relatively weak (flux levels a few percent of the continuum level, but statistically significant detections), and very broad (on the order of 1000s of \kms), confirming they are formed in the stellar wind. The profiles have a flat-topped morphology, suggesting they are formed in an optically thin shell of material. The high energies (55, 97, and 126 eV, respectively) and low critical densities (log($n_e$/cm$^{-3}$) = 4.2, 5.2, and 6.0, respectively) \added{required to reach these ionization stages} also indicate the presence of a very hot, tenuous wind component. These characteristics make the MIR lines an exciting new tool for studying hot, rarefied components of stellar winds. However, concerns were raised on the broader applicability of these diagnostics as the lines could not be detected in a comparable MIR spectrum of $\mu$~Col, an O9.5~V star with very similar stellar properties to 10~Lac. \added{MIR fine structure lines are however a well established feature in Wolf-Rayet stars (e.g. \citealp{Barlow1988, Dessart2000}). Generally, the lines observed are relatively low ionization species of Ne and S, but do extend to \ofour\ 25.9\micron\ (e.g. \citealp{Morris2004})}. It was therefore unclear whether MIR fine-structure lines would be a new diagnostic tool for O-type stars or a rare phenomenon only arising under very specific conditions. 

Fortunately, a larger sample of OB-type stars was soon observed in the MIR with JWST as part of dust extinction curve surveys \citep{decleir25,zeegers25}, offering the opportunity for a preliminary investigation on the incidence rate of fine-structure wind emission in massive stars. In this work, we extend the efforts of \cite{Law2024} to all OB stars observed with JWST-MIRI up to JWST Cycle 4 and discuss the viability of these lines as diagnostics for wind physics across the parameter space covered. 

We describe the observations and data processing techniques in \S \ref{obs.sec}, present results on line detections and wind parameter estimates in \S \ref{results.sec}, discuss the implications of these results in \S \ref{discussion.sec}, and conclude in \S \ref{conclusions.sec}.


\begin{figure*}[!htbp]
\epsscale{1.2}
\plotone{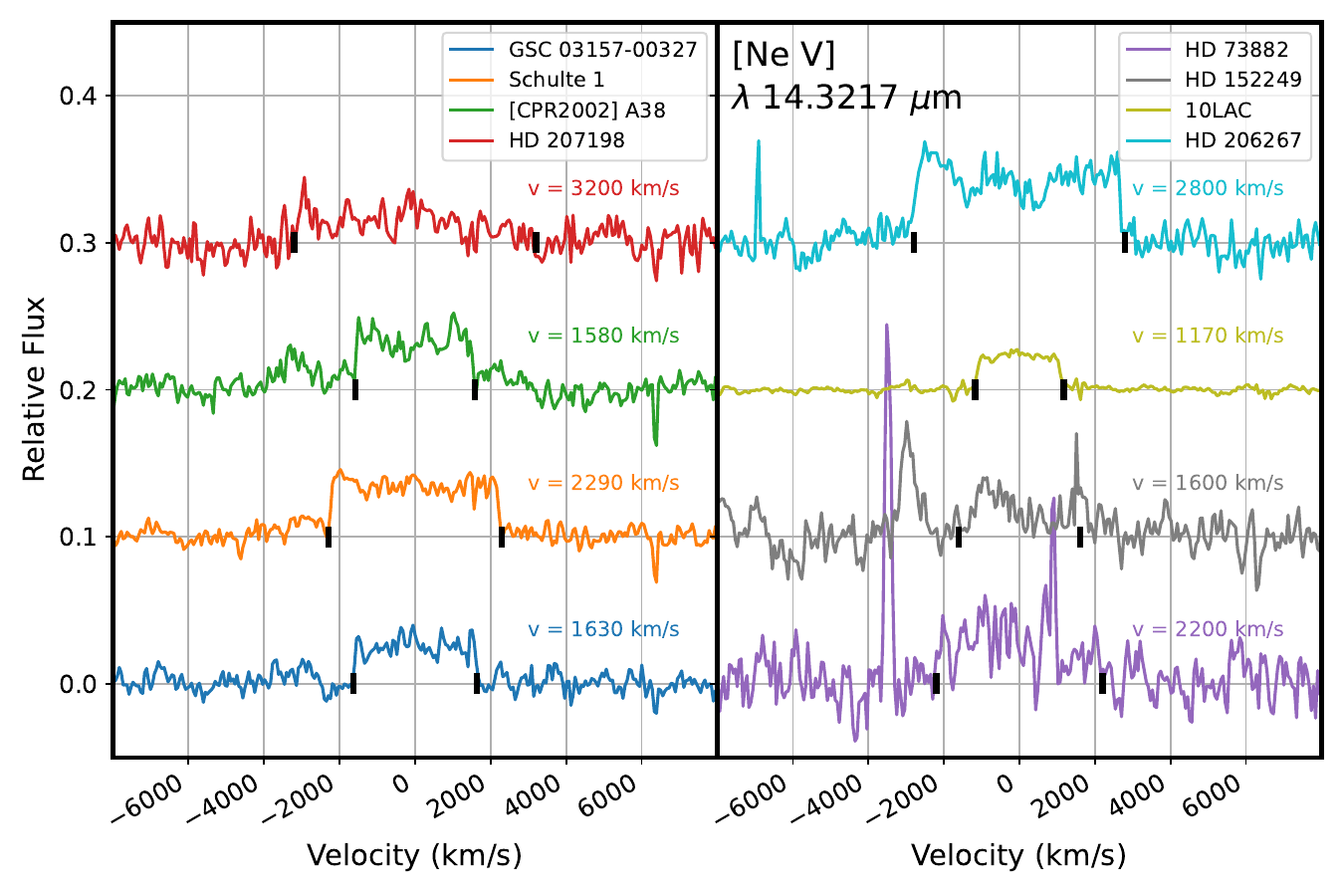}
\caption{Continuum-subtracted spectra for all stars in which statistically significant \nev\ emission is detected.  Each spectrum has been continuum-subtracted and normalized by the median continuum value at 14.3 \micron; a value of 0.1 thus represents a strength 10\% of the stellar continuum.  Note that each spectrum has been offset by 0.1 for clarity; the typical \nev\ line strength peaks at about $2-5$\% of the continuum.  Wavelengths are given in velocity units relative to the 14.3217 $\micron$ rest wavelength of the \nev\ emission feature, vertical black lines atop each spectrum denote the maximal wind velocities derived in \S \ref{results-vinf.sec} and denote the integration limits to measure EWs.
The feature at $-$2900 \kms\ is H\one\ 14.183 $\micron$ (13--9). Note that GSC 03157-00327 is referred to as ALS 15134 throughout the text. 
}
\label{spectra.fig}
\end{figure*}

\section{Observations and Data Processing}
\label{obs.sec}

We compiled a sample of 22 massive stars drawn from a combination of JWST programs: the Webb Investigation of Silicates, Carbons, and Ices program (WISCI; 11 stars; PID 2183, PI Zeegers) and the Measuring Extinction and Abundances of Dust program (MEAD; 9 stars; PID 2459, PI Decleir), augmented by calibration observations of 10 Lac (PID 1524; PI Law) and $\mu$ Col (PID 4497; PI Gordon). The WISCI and MEAD program stars were originally chosen for their location along sightlines with sufficient dust optical depth to study weak absorption features in the diffuse ISM of the Milky Way \citep{zeegers25, decleir25}, and therefore contain an assortment of 10 B-type stars, 11 late O-type stars and 1 earlier O-type star. We omit the O8 type WISCI source LS 4992 from our sample since it has an extremely rich spectrum dominated by dust features along the line of sight that severely complicates our search for faint, broad emission features in the spectrum of the star itself.

All 22 stars were observed using the MIRI Medium Resolution Spectrometer (MRS) aboard JWST across the full wavelength range from 5--26.5 $\micron$ with a maximum resolving power $R \sim 4000$ at 5 $\micron$ that decreases to $R \sim 1500$ at 25 $\micron$ \citep{Argyriou2023}. \added{For the lines discussed in this work, \nevi\ 7.65 \micron, \nev\ 14.32 \micron, \nev\ 24.89 \micron\ and [O\four] 25.89 \micron\ the resolving power is $R = 3600, 2800, 1500$ and $1300$, respectively.} Since these stars were observed as a combination of calibration programs and probes of the dusty interstellar medium, the observations span a wide range of achieved continuum signal-to-noise ratios (S/N) which are reported in Table \ref{tbl:targets}.

We reduced the data using the JWST calibration pipeline version 1.20.2\footnote{This provides an updated flux calibration relative to \citet{Law2024} that fixes narrow spectral artifacts seen in their analysis of the \nev\ emission feature, and improves the calibration in the vicinity of the \nevi\ emission feature.} and the publicly-available pipeline notebooks provided by \citet{law_zenodo}.  We used the default pipeline settings, except for enabling the optional correction for residual cosmic ray showers and using a smaller extraction radius of 1 FWHM (compared to the default of 2 FWHM) in order to maximize the SNR of the stellar spectra at the expense of marginally worse relative flux calibration between adjacent spectral bands. We chose to use the 1-d residual fringe corrected spectra provided by the pipeline since there are no strong molecular bands in our targets.
Additionally, for 10 Lac we note residual periodic fixed-pattern noise in the vicinity of \nev\ that was not removed by the residual fringe correction; we estimate the magnitude and phase of this noise from the continuum region and subtract it from the \nev\ wavelengths.

\begin{deluxetable*}{ccccccccccc}
\tabletypesize{\small}
\label{tbl:targets}
\tablecolumns{11}
\tablecaption{\nev\ 14.3\micron\ Detections with Continuum S/N for Target Stars}
\tablehead{
\colhead{Source} & \colhead{Spectral Type} & \colhead{Detection} &
\colhead{Flux} & \colhead{EW} &
\colhead{S/N at 14\micron} & \colhead{S/N at 7\micron} & \colhead{S/N at 26\micron} &
\colhead{PID} & \colhead{Obs}
}
\startdata
HD 206267\tablenotemark{a}       & O5 V     & yes & $28.0 \pm 0.6$  & $114 \pm 3$    & 118 & 183 & 4   & 2459 & 14 \\
ALS 15134                  & O8 Vz     & yes & $0.78 \pm 0.03$ & $40 \pm 1$     & 151 & 188 & 6   & 2183 & 10 \\
{[}CPR2002{]} A38                & O8 V     & yes & $0.62 \pm 0.02$ & $48 \pm 1$     & 143 & 184 & 5   & 2183 & 4  \\
Schulte 1       & O8 V + B     & yes & $2.74 \pm 0.05$ & $72 \pm 1$     & 162 & 205 & 8   & 2183 & 1  \\
10 Lac                           & O9 V     & yes & $3.75 \pm 0.06$ & $25.0 \pm 0.4$ & 479 & 391 & 95  & 1524 & 16 \\
\hline
HD 207198                        & O8.5 II  & ?   & $9.7 \pm 0.6$   & $42 \pm 3$     & 108 & 207 & 4   & 2459 & 16 \\
HD 73882                         & O8.5 IV  & ?   & $6.6 \pm 0.6$   & $64 \pm 6$     & 64  & 120 & 0   & 2459 & 6  \\
HD 152249                        & OC9 Iab     & ?   & $4.7 \pm 0.5$   & $27 \pm 3$     & 81  & 137 & 3   & 2459 & 10 \\
\hline
$\mu$ Col                        & O9.5 V   & no  & - & -     & 236 & 218 & 26  & 4497 & 4  \\
ALS 15181                        & B0 V   & no  & - & - & 90  & 172 & 5   & 2183 & 7  \\
TYC 7380-1046-1                  & B0 Ia     & no  & - & -  & 159 & 166 & 14  & 2183 & 19 \\
TYC 8989-436-1                   & B0.5 II  & no  & - & - & 173 & 191 & 10  & 2183 & 35 \\
HD 203938                        & B0.5 IV  & no  & - & - & 73  & 156 & 2   & 2459 & 12 \\
GSC 08152-02121                  & B2 IV    & no  & - & - & 113 & 170 & 6   & 2183 & 32 \\
2MASS J20452110+4223513          & B2 V     & no  & - & - & 149 & 195 & 7   & 2183 & 13 \\
TYC 6272-339-1                   & B3 II    & no  & - & - & 153 & 207 & 7   & 2183 & 22 \\
CPD-59 5831                      & B5 Ia     & no  &  - & - & 139 & 221 & 13  & 2183 & 38 \\
CD-40 11169                      & B8 Ia     & no  &  - & -  & 170 & 217 & 10  & 2183 & 16 \\
\hline
HD 14434                        & O5.5 V   & -   & - & - & 23  & 42  & 0   & 2459 & 2  \\
HD 216898                       & O9 V     & -   & - & -  & 31  & 96  & 0   & 2459 & 18 \\
HD 38087                         & B3 II   & -  & - & - & 14  & 69  & 0   & 2459 & 4  \\
HD 147888                       & B3 V     & -   & - & - & 32  & 199 & 3   & 2459 & 8  \\
\enddata
\tablenotetext{}{Spectral types are taken from \cite{zeegers25} and \cite{decleir25}.}
\tablenotetext{}{S/N columns refer to continuum S/N at the respective wavelengths.}
\tablenotetext{a}{O5V((fc)) + B0:V + O9V \citep{MaizApellaniz2020}, O6V + O9V \citep{decleir25}.}
\end{deluxetable*}


\section{Results}
\label{results.sec}

\subsection{MIR Wind Detections}
\label{results-detect.sec}

For each source, we fit the stellar continuum in the flux-calibrated spectrum 
provided by the pipeline using a 3rd-order polynomial in the wavelength ranges $\lambda\lambda 13.6 - 14.1$ \micron\ and $\lambda\lambda 14.5 - 15.0$ \micron\ (i.e., omitting the wavelength region within 4000 \kms of \nev).
Figures \ref{spectra.fig} and \ref{nondet.fig} plot the resulting continuum-subtracted spectra of each of our 22 targets in the vicinity of the \nev\ 14.3217 \micron\ spectral feature; eight targets show evidence of broad \nev\ emission.

In Table \ref{tbl:targets} we give the S/N of the spectral features using an empirical noise estimate based on the rms of the continuum-subtracted spectrum adjacent to the \nev\ spectral feature and near the rest wavelengths of \nevi\ 7.6524 \micron\ and [O\four] $25.89$ \micron.

\begin{figure}[tbp]
\epsscale{1.2}
\plotone{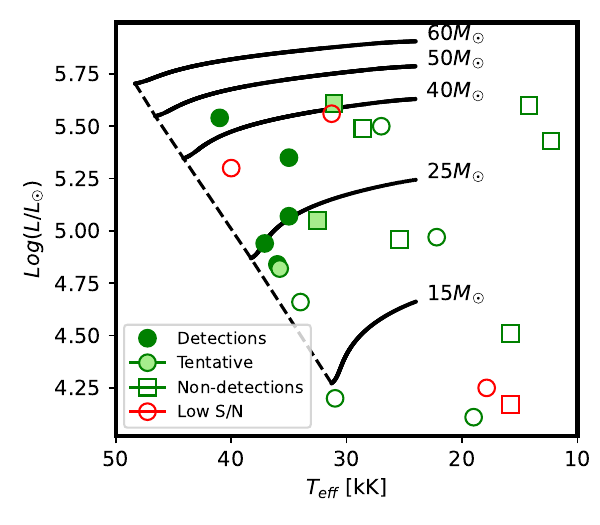}
\caption{HRD showing the full sample discussed in this work. Symbols in green have sufficient S/N to allow us to make some comment on the nature of the line detection, those in red have very poor S/N. Filled symbols indicate a detection in \nev\ 14.3\micron, those filled with a lighter color are the tentative detections and open symbols are non-detections. Shapes indicate the luminosity class, circles for dwarf stars and squares for (super)giants. Black lines are \added{main sequence} evolutionary tracks for non-rotating Galactic massive stars from \cite{Brott2011}, labeled with their respective initial masses. The dashed line indicates the zero-age main sequence.
}
\label{HRD.fig}
\end{figure}

We quantify the status of the \nev\ 14.3217 \micron\ detections in Table \ref{tbl:targets} and show this distribution in Fig. \ref{HRD.fig}. For 4 stars (O5.5~V, O9~V, B3~II and B3~V) the S/N is too low to comment on the presence of the emission lines. \nev\ emission that exceeds the scatter in the local continuum by a factor of 3 or more is found in 8 stars, 5 of which are strong detections and 3 of which are more tentative. Of the 5 strong detections, 4 are classified as late O dwarfs (O8~V, O8~V, O8~V and O9`V) and 1 is a multiple system containing an inner O5~V + B0:V binary with a wider O9~V companion (with the MIR emission likely associated with the O5~V star, discussed in Sect. \ref{results-vinf.sec} and Appendix \ref{notes.sec}). The 3 tentative detections are also late spectral types but higher luminosity classes (O8.5~IV, O8.5~II and O9~I). Emission features are not detected in the rest of the sample, which consists of two O9.5~V stars and 9 B-type stars with a wide range of spectral types and luminosity classes. 

\begin{figure}[tbp]
\epsscale{1.2}
\plotone{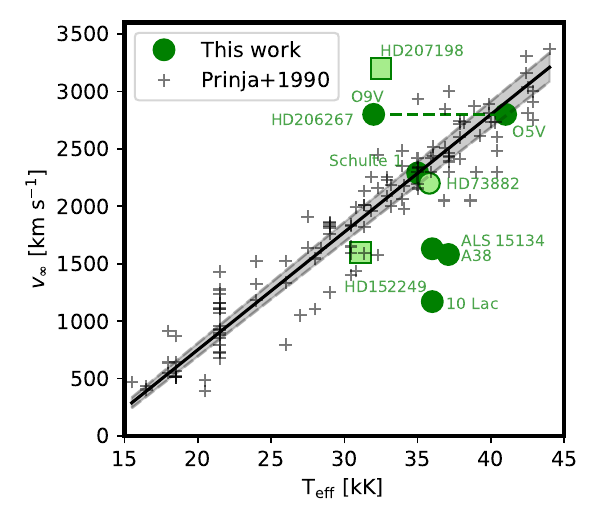}
\caption{Terminal wind speeds as a function of effective temperature for the 5 stars with strong \nev\ detections \added{(filled dark green circles) and 3 stars with tentative detections (filled light green circle and squares, as in Fig. \ref{HRD.fig}])}. Green points correspond to measurements from the \nev\ 14.3 \micron\ line from this work, the gray sample is terminal wind speeds measured for Galactic OB stars from UV spectroscopy in \cite{Prinja1990}. Note the wind speed of HD~206267 is shown for two temperatures, to illustrate the positions if the emission is arising from the O5~V or O9~V components of the multiple system. 
}
\label{vinf_pred.fig}
\end{figure}

Repeating the above analysis at other wavelengths, we note that most of these eight stars also show evidence for \nevi\ 7.6524 $\micron$ emission (Figure \ref{nevi.fig})
similar to that observed in 10 Lac \citep{Law2024}, while \nev\ 24.31 \micron\ and \ofour\ 25.89 $\micron$ emission is more difficult to confirm (e.g. Fig. \ref{oiv.fig} for \ofour). Generally, the \nevi\ spectral width is consistent with the observed \nev\ width.  However, given the low S/N of the WISCI and MEAD observations at these wavelengths, it is difficult to characterize these fainter line features in detail and we therefore focus our attention in the present contribution exclusively on the \nev\ 14.32 $\micron$ emission feature.


The MIRI wavelength range also covers fine structure transitions of {[Ne\two}] 12.81 $\micron$ and [Ne\three] 10.86, 15.56 $\micron$. These features are not seen in any of our sample with the exception of HD~206267 where [Ne\three] 15.56 $\micron$ is detected. As discussed in the Appendix, this object is complex with at least 2 components. It is not clear which component produces the [Ne\five] and [Ne\three] emission but the widths of the features suggest it is the O5~V star component.


%
\begin{figure}[tbp]
\epsscale{1.2}
\plotone{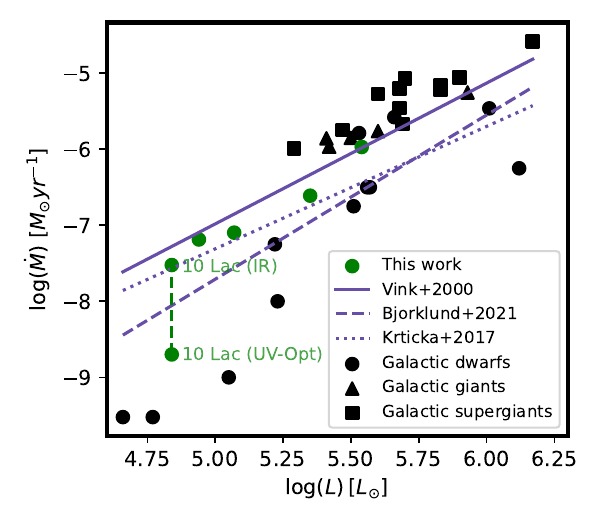}
\caption{Mass-loss rates as a function of luminosity for the 5 stars with strong \nev\ detections, shown as green circles. Black markers show a sample of Galactic O-type stars from \cite{Repolust2004, Martins2005b, Crowther2006} for comparison. Circles, triangles and squares correspond to luminosity classes V, III and I respectively. Solid and dashed lines show theoretical mass-loss rate predictions from \cite{Vink2000} and \cite{Bjorklund2021} respectively. 
}
\label{lum-mdot.fig}
\end{figure}

\subsection{Wind Speeds}
\label{results-vinf.sec}
As the \nev\ 14.3 \micron\ line is formed in an optically thin shell of expanding material, as evidenced by the flat-topped profile morphology, the terminal wind velocity ($v_{\infty}$) can be estimated by measuring the half-width of the profile at zero intensity in continuum-subtracted spectra. We compare the measured wind speeds to empirical predictions from \cite{Hawcroft2023}. The relation provided in \cite{Hawcroft2023} is based on the \cite{Prinja1990} sample for Galactic OB stars, shown in Fig \ref{vinf_pred.fig}. Of the 5 stars with visibly flat-topped profiles, the two stars with stronger winds tend to agree more closely with the general empirical trend: Schulte 1 lies well within the predicted wind speed region; along with HD~206267, assuming the \nev\ emission is coming from the O5~V component (as opposed to the O9~V binary companion). 
The stars with relatively weaker emission (A38 and ALS 15134) lie significantly below the empirical prediction, by $\sim 700$ \kms. 10 Lac has the largest discrepancy compared to the empirical prediction, by $\sim 1000$ \kms. 

\begin{figure}[tbp]
\epsscale{1.2}
\plotone{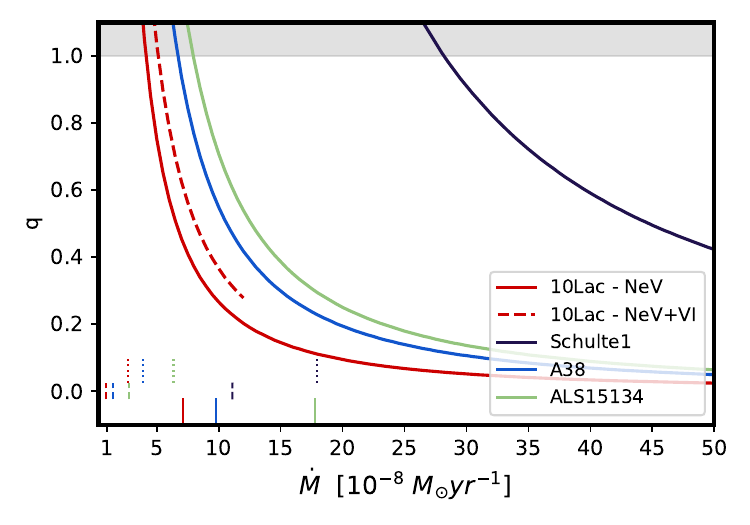}
\plotone{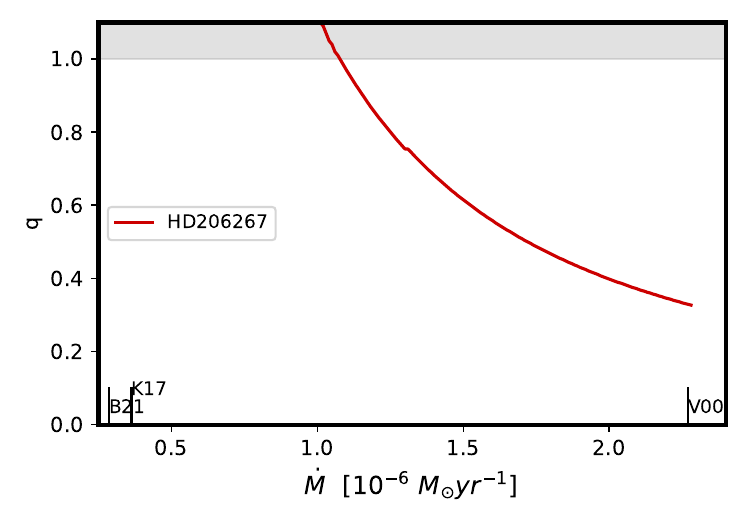}
\caption{Ionization fraction $q$ as a function of mass-loss rate for Ne$^{4+}$. All curves are with an electron temperature of $T_{e} = $ 35kK. Vertical ticks show theoretical mass-loss rate predictions from \cite{Bjorklund2021}, dashed, \cite{Krticka2017}, dotted and \cite{Vink2000}, solid.
}
\label{mdot-q.fig}
\end{figure}

The three stars with more tentative detections do not have clear flat-topped morphologies and so any estimates of \vinf\ are less robust. \added{HD~207198 sits well above the empirical prediction when estimating the width of the statistically significant emission, but given the significance of the detection is already low it would not be surprising if this is an overestimate. From a visual inspection of the spectrum as shown in Fig. \ref{spectra.fig} it would be reasonable to estimate a \vinf\ of $\sim2000$\kms\ which is closer to the empirical prediction}. HD~73882 and HD~152249 do however lie along the observed trend.

\subsection{Mass-loss rates}
\label{results-mdot.sec}
Using the techniques described in \citet{Law2024}, and defined in \citet{Barlow1988, Dessart2000, Crowther2024}, we estimate the mass-loss rates required to drive the observed line flux in \nev 14.3 \micron. Using these relations, \added{specifically Equation (A3) from \cite{Law2024}}, we inherit the assumption that the upper levels are populated only through collisional excitation, however contributions from UV continuum fluorescence or non-radiative processes could be significant. \added{We also adopt assumptions from \cite{Law2024} on the mean ionic mass and mean number of electrons per ion, commonly $\sim1.5$ and $\sim0.8$ respectively for O-type stars. We assume a solar neon abundance of log(Ne/H)+12 = 8.09 for all stars apart from 10~Lac, for which we use the photospheric abundance measured in \cite{Aschenbrenner2023} of log(Ne/H)+12 = 8.24. An increase in Ne abundance would lower the mass-loss rate, further reducing the lower limits estimated here, however the shift is relatively small. For example in the case of 10~Lac, an increase of 0.2 dex in Ne would result in a downward shift of 0.1 dex in log\Mdot. Finally, we assume a smooth wind for the emission region, in line with the \cite{Lucy2012} scenario where high-ionisation emission arises from the hot, rarefied interclump wind, meaning the emission lines measured here should not be affected by enhanced emission arising from cooler, overdense clumps of material.} The ionization fraction \added{($q$, the fraction of atoms in a given ionization state, in this case \textrm{Ne\,{\sc v}})} as a function of mass-loss rate is shown for the stars with strong \nev\ 14.3~\micron\ detections in Fig. \ref{mdot-q.fig}. The estimates of limits on mass-loss rates should be considered with the caveat of only collisional excitation in mind, until these results can be tested with more detailed stellar atmosphere models, the development of which is beyond the scope of this work. Additionally, we assume \nev\ contributes the full ionization fraction and use an electron temperature $T_e =$ 35\,kK to provide lower limits on the observed mass-loss rate, any reduction to the ionization fraction of \nev\ or increase in electron temperature would increase the required mass-loss rate. \added{An example of the increased mass-loss rate with $q$ relation as a result of increasing electron temperature to $T_e =$200\,kK for 10~Lac is shown in Figure 4 of \cite{Law2024}.}

As shown in Fig. \ref{lum-mdot.fig} we observe a trend of increasing mass-loss rate with luminosity. \added{The lower limits on mass-loss rates for the sample all lie within the boundaries of theoretical mass-loss rate predictions. We find lower limits slightly above the \cite{Bjorklund2021} and \cite{Krticka2017} predictions but below the \cite{Vink2000} rates. This is in good agreement with generally observed trends between empirical and theoretical mass-loss rates, which come from studies based on UV/optical spectroscopy of typical O-type stars that take into account the effects of clumping on the observed line profiles }(e.g. \citealp{Hawcroft2021, Brands2022, Backs2024, Hawcroft2024}). If the limits found here are good representations of the true mass-loss rates, the best agreement would be with the \cite{Krticka2017} rates, which under-predict the observed mass-loss rate lower limit by a factor of 1.7 on average, while \cite{Vink2000} and \cite{Bjorklund2021} over- and under-predict the observed values by factors of 3.5 and 2.6, respectively. 

\subsection{Profiles of Fine-Structure Lines}
\label{results-profiles.sec}

\begin{figure*}[!htbp]
\epsscale{1.0}
\plotone{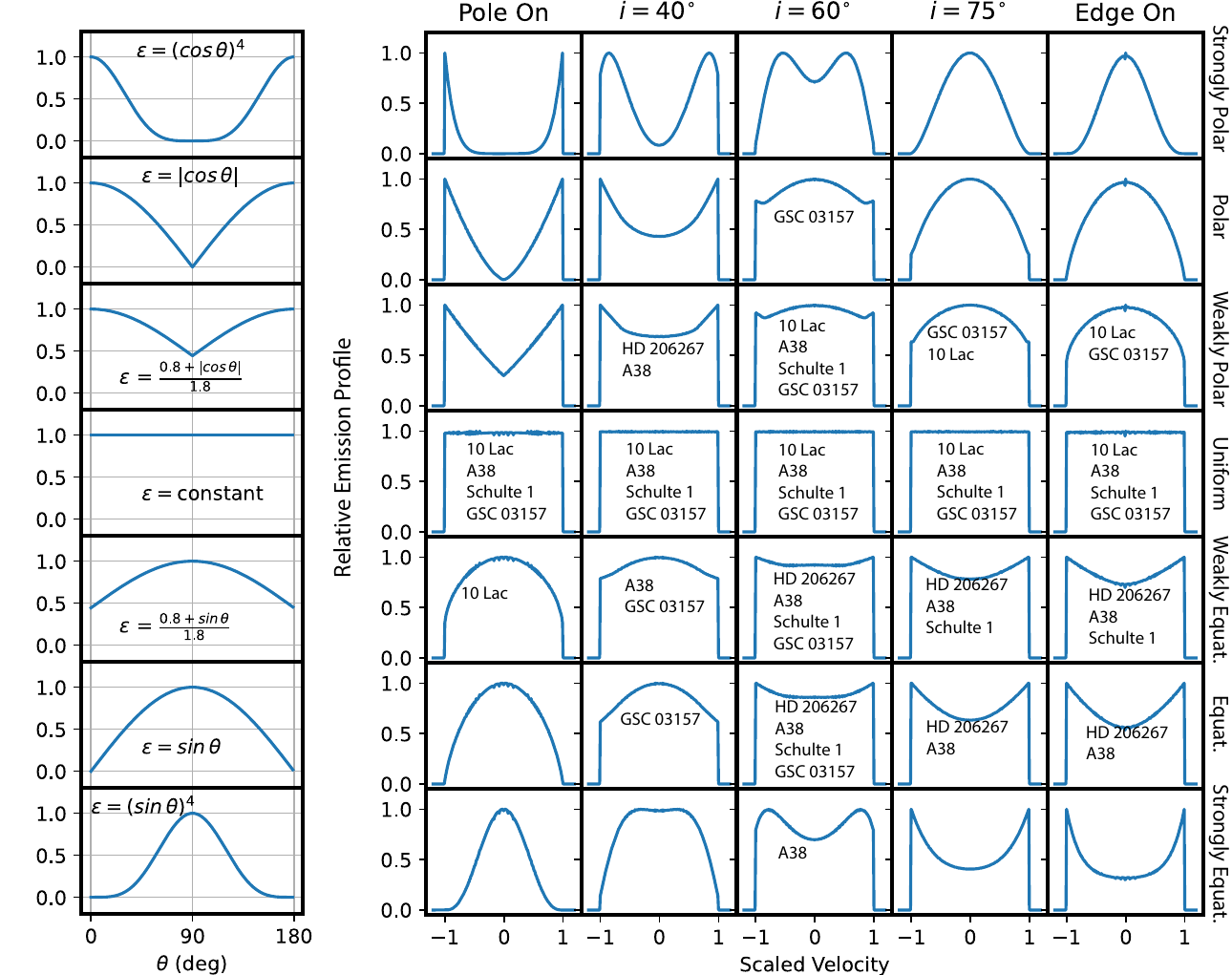}
\caption{Left panel: Density profiles $G(\theta)$ as a function of the polar angle $\theta$.  
         Right panel: Continuum-subtracted profiles of fine-structure lines that would be seen by a distant observer as a function of the viewing inclination $i$.  
         Labels within each panel list the stars whose spectra are statistically consistent with 
         the computed profile.
        }
\label{geometry.fig}
\end{figure*}

Figure~\ref{spectra.fig} indicates that the shape of the {\nev} 14.32 \micron\ line is usually consistent at the $2\sigma$ level with the square, flat-topped profile expected  for optically thin emission in a spherically symmetric outflow (e.g. \citealp{ignace06}). However, significant deviations from a flat-top profile also occur, in particular for HD~206267, which exhibits double-peaked ``horns'' at its wavelength extrema. The profile of {\ofour}\,24.89 \micron\ in the spectrum of 10~Lac has a similar morphology \citep{Law2024}.

We explored possible origins for the double-horned morphology by computing line profiles for distributions of outflowing material that deviate from spherical symmetry in stellar co-latitude while maintaining axi-symmetry. Our simplified approach and its limitations are described in Appendix~\ref{AppendixC}. Figure~\ref{geometry.fig} shows a selection of profiles for distributions of material that are not based on specific physical models, but are only intended to illustrate the effects of non-spherical distributions of density and the aspect of the viewer. As expected, a purely radial distribution of density produces a flat-topped line profile at all viewing angles. However, as previously shown by \cite{ignace06}, distributions that are enhanced at the stellar  pole or equator produce line profiles that range from double-horned to rounded and centrally peaked. For example, double-peaked profiles are produced both by strongly polar emission seen pole-on and by strongly equatorial emission seen edge-on. For a given distribution of material, the profile depends strongly on the inclination angle of the observer.

Although the double-horn morphology appears to be a clear signature of an aspherical distribution of material, a more detailed interpretation requires additional information. In the case of HD~206267, Figure~\ref{geometry.fig} suggests that the double-horn profile results from a modest enhancement of material at the stellar equator viewed by an observer inclined to the rotation axis by $60^\circ-90^\circ$. In this connection, it is interesting to note that \citet{raucq18} estimated the inclination of the orbital plane of HD~206267 to be $i = 76^{\circ}$ by comparing spectroscopic and dynamical masses. If the normal to the orbital plane is aligned with the rotation axis, the similarity in viewer aspect suggests the existence of enhanced density in the vicinity of the orbital plane of this complex system.

\section{Discussion}
\label{discussion.sec}

\subsection{Wind Speeds}
\label{disc-vinf.sec}



\begin{table}
\centering
\caption{Wind Speeds for stars with \nev\ detections.}
\label{windspeeds.table}
\small
\begin{tabular}{ccc}
\hline
ID & \vinf\ (UV) & \vinf\ (MIR) \\
   & [\kms]       & [\kms]      \\
\hline
HD 206267     & 2815$^{a}$ & 2800 \\
ALS 15134     & --         & 1630 \\
{[}CPR2002{]} A38 & --     & 1580 \\
Schulte 1     & --         & 2290 \\
10~Lac        & 1070$^{b}$ & 1170 \\
HD 207198     & 2090$^{c}$ & 3200$^{d}$ \\
HD 73882      & 2315$^{c}$ & 2200 \\
HD 152249     & 2100$^{c}$ & 1600 \\
\hline
\end{tabular}

\vspace{2mm}

{\footnotesize
$^{a}$ Measured directly from STIS spectra. \\
$^{b}$ From \cite{Martins2005b}. \\
$^{c}$ From \cite{Prinja1990}. \\
$^{d}$ Measurement from statistically significant emission, likely an overestimate from visual inspection.%
}
\end{table}
To first attempt to verify whether the outliers in wind speed measurements are related only with the MIR measurements, or whether UV diagnostics also often result in outliers from the general trend, we compare with empirical calibrations drawn from a large sample of Galactic OB stars \citep{Prinja1990}. UV diagnostics do not seem to result in outliers in wind speed, as the \cite{Prinja1990} sample does not contain a sub-sample of stars with $T_{\rm{eff}}>30$kK, that show saturated UV P-Cygni profiles and have significant outlier measurements of \vinf\ (especially after updating the stellar parameters with the latest literature, see Appendix \ref{notes.sec} for details). 

The temperatures for the 3 stars in this sample with discrepant \vinf\ measurements (A38, ALS 15134 and 10~Lac), have all been measured with fitting of synthetic spectra from tailored stellar atmosphere models (for A38 and ALS 15134 see \citealp{Berlanas2020} and for 10~Lac see \citealp{Aschenbrenner2023}), making them robust to within a few kK, and ruling out the possibility that they could each be $>5$\,kK cooler, which would be required to agree with the empirical predictions. This supports the notion that the wind speeds measured from the MIR for these 3 stars are genuine outliers from the general trend. It is unclear whether the wind speeds, in that case, would be underestimated due to the MIR lines not tracing the true terminal wind speed of the global outflow, or whether the wind speeds are indeed significantly lower than predicted. Similarly, these 3 stars have low wind speed to escape speed ratios (\vinf /$v_{esc}$ = 1.1, 1.8 and 1.6 for 10~Lac, ALS 15134 and A38 respectively) while the other stars sit close to the \vinf /$v_{esc}$ relation measured in \cite{Hawcroft2023} (see Fig. \ref{vesc_pred.fig}).

As comparison points for consistency in wind speeds \added{we list wind speeds measured from the UV and MIR in Table \ref{windspeeds.table}}. Unfortunately, HD~206267 (O5~V) is the only object with strong wind lines in the UV and MIR. We can also consider HD~207198, HD~152249 and HD~73882 which have strong UV lines and tentative detections in the MIR. HD~207198 and HD~152249 have visually very weak MIR detections and estimates of $v_{\infty}$ from the MIR, which do not agree with those measured in the UV. HD~73882 has a slightly more visible signature in the MIR and the estimate of $v_{\infty}$ is in agreement with that measured from the UV. It is therefore difficult to say from these 3 stars whether the UV and MIR give consistent wind parameters. For HD~206267, the wind speed can be robustly measured from the \ion{C}{4} $\lambda$ 1548~\AA\ and \ion{N}{5} $\lambda$ 1240~\AA\ lines as well as the \nev\ 14.3\micron\ line. The estimates on $v_{\infty}$ from the UV are consistent (within 110 \kms) with those measured from the \nev\ line and are discussed further in Appendix \ref{notes.sec}. HD~206267 gives the preliminary impression that wind speeds from the hot component visible in the MIR are consistent with the cooler wind component traced by UV resonance lines, at least for a relatively early (O5~V) O-type star. However, given the limitations on the scope of the current sample, and a lack of options for multi-wavelength follow-up on targets with high visual extinction, it is difficult to draw further conclusions on the wind speeds of these objects. 

Nonetheless, with confirmation of the MIR lines across a wider sample, we have a new window to measure wind speeds for the first time at longer wavelengths. Such information was only previously available in the UV. These MIR lines are also viable wind diagnostics for stars with either no strong UV wind signatures or high extinction. Since they probe different density regions, the simultaneous presence of UV and MIR wind diagnostics may provide additional constraints on the velocity structure of the wind.

\subsection{Mass-loss rates}
\label{disc-mdot.sec}



The long-standing weak-wind problem is one of the most pressing open questions related to stellar winds, and may be the largest source of uncertainty in wind strengths for O-type stars on the main sequence (e.g. \citealp{Puls2008, Vink2022}). Recent efforts exploring mass-loss rate relations with key parameters like structure and metallicity have greatly reduced uncertainties on wind strengths over the majority of the main sequence. However, even for typical O-type stars at extreme ends of the luminosity range $ 5.2 < \log(L/L_{\odot} ) < 6.0$ the magnitude and nature of the wind remains unclear. Focusing on the low-luminosity weak-wind stars, the issue refers to a sudden reduction in strength of typical wind diagnostics in the UV-optical regime in stars with $\log(L/L_{\odot} ) <5.2$, resulting in a drop in inferred mass-loss rates by around 2 orders of magnitude \citep{Bouret2003}. This reduction is a strong discontinuity from the general observed trend and has been shown to be at odds with established theoretical predictions across a range of environments with varying metallicities \citep{Martins2004, Martins2005, Marcolino2009, Hawcroft2024, deAlmeida2019}. 

There are two main avenues towards a solution, either a true reduction in wind strength or maintaining wind strength but shifting away from typical diagnostics, both of which highlight missing physics in our current wind models. Multiple theoretical solutions have been proposed in both cases. For the latter, \cite{Lucy2012} envisioned a scenario where the low density of the winds in late O-type stars results in inefficient cooling relative to the much denser winds in early O-type stars. In combination with shock-heating due to the inherent line-deshadowing instability, this suggests that the ionization state would be too high to form wind line profiles in the typical UV diagnostics and would shift to higher ionization transitions. This scenario has been further explored in hydrodynamic simulations of O-type star atmospheres by \cite{Lagae2021}. 
In the alternate case, that winds truly are weaker, \cite{Vink2022} suggests a reduction in line-driving force could arise from the recombination of iron to a lower state with fewer transitions. \cite{Lattimer2021} provide an updated formalism for radiative line driving and predict a sharp decrease in wind strength below 15\,kK. While this threshold temperature is too low to fully account for the O-type star weak-wind problem, it identifies a mechanism which could partially contribute and highlights that line driving formalisms could be revised. \cite{Owocki2002} discuss that a reduction in line force could arise from the decoupling of minor ions from the bulk wind. 

To test and distinguish between the proposed scenarios, a few alternate observational diagnostics have been highlighted in the literature, such as near-infrared hydrogen lines which are known to be more sensitive to changes in mass-loss rates at low values \citep{Najarro2011}. Although estimates of mass-loss rates using Br$\alpha$ in \cite{Garcia2025} agree with values found from UV and optical diagnostics. Individual cases of higher mass-loss rates have been observed through extreme X-ray flux in $\mu$~Col \citep{Huenemoerder2012}, a star with very weak UV and optical wind lines, which supports the \citet{Lucy2012} scenario. Further evidence in this direction comes from the detection of high ionization fine-structure line emission in the mid-infrared in 10~Lacertae \citep{Law2024}. Although, curiously $\mu$~Col shows no trace of wind emission in the MIR.

The detections of high ionization emission in the MIR in 4 additional stars presented here builds on the 10~Lac finding. The preliminary agreement between the observed and predicted mass-loss rates for these stars in the weak-wind regime using the \nev 14.3\micron\ line is enticing, and there is clearly an avenue to measure mass-loss rates from this line (and other high ionization fine structure emission lines in the MIR). However, we cannot yet place strong constraints on the mass-loss rates given the caveats surrounding our assumption of collisional excitation built into the relation used to make our estimates. As a result, we delay further discussion to future work which can be informed by more detailed, tailored stellar atmosphere models.

A comparison between multi-wavelength mass loss diagnostics for this sample would also be informative. Unfortunately we cannot make such comparisons for any stars aside from 10~Lac as there is no mass-loss information available in the literature from studies at short-wavelengths. There are no archival UV spectra for ALS 15134 or A38 and there are flux gaps in the STIS UV spectra of Schulte 1 over the CIV $\lambda$ 1548\AA\ line. There are existing STIS optical spectra, although with insufficient S/N and resolution for detailed stellar atmosphere modeling. The large visual extinction for these targets ($A_{V} > 5$) limits the possibility for follow-up spectroscopic observations and analysis at shorter wavelengths. For HD~207198, HD~73882 and HD~152249 there are extensive UV observations with strong wind lines, however the only tentative detections of emission in \nev\ again limit a multi-wavelength comparison. For HD~206267 the extinction is less extreme and there are archival UV and optical spectra. In this case we can make a comparison with the terminal wind speeds as the UV P-Cygni profiles are saturated, but the UV spectra have not been modeled in detailed with stellar atmosphere codes in the literature, likely due to the additional complications of this being a multiple system including a close binary. While it may be possible to disentangle the components to obtain a clean spectrum of each component using existing observations/tools (discussed further in Appendix \ref{notes.sec}), such efforts are beyond the scope of this work. 


\begin{deluxetable*}{ccccccccccccc}
\tabletypesize{\scriptsize}

\label{tbl:params}
\tablecolumns{7}
\tablecaption{Stellar parameters for stars with strong \nev\ detections}
\tablehead{
\colhead{ID} & \colhead{SpT} & \colhead{$v_{\infty}$} & \colhead{log($\dot{M}$)} & \colhead{log$(\frac{L}{L_{\odot}})$} & \colhead{$T_{\rm{eff}}$} & \colhead{log\,$g$} & \colhead{$R$}  & \colhead{$v_{\infty}$ [pred]} & \colhead{$v$ sin $i$} & \colhead{$N_{He}/N_{H}$} & \colhead{log($\dot{M}$)[lit]} & \colhead{log$L_{X}/L_{bol}$} \\
\colhead{} & \colhead{} & \colhead{[\kms]} & \colhead{[dex]} & \colhead{[dex]} & \colhead{[kK]} & \colhead{[dex]} &\colhead{[$R_{\odot}$]} & \colhead{[\kms]} & \colhead{[\kms]} & \colhead{} & \colhead{[$M_{\odot}yr^{-1}$]} & \colhead{[dex]}
}
\startdata
ALS 15134 &  O8 Vz & 1630 & -7.1 & 5.07\tablenotemark{a} &	36	& 3.8 & 8.8  & 2270 & 95 & 0.099 & -8.5\tablenotemark{b} & -7.1\tablenotemark{b} \\
{[}CPR2002{]} A38  & O8 V  & 1580  & -7.2  & 4.94\tablenotemark{a, c}	&	37.1	& 4.03	& 7.2 & 2370 & 120 & 0.105 & - & -\\
Schulte 1 & O8 V + B & 2290 & -6.6 & 5.35\tablenotemark{i}	& 35 & 3.69 & 12.8 & 2180	& 185	& $<$0.066 & & \\
HD 206267  & O5 V & 2800 & -6.0 & 5.49\tablenotemark{d} & 40.9 & 3.9 & 11.2 & 2730 & ? & ? & ? & -6.47\tablenotemark{j} \\
10~Lac  & O9 V & 1170 & -7.5 & 4.84\tablenotemark{h} & 36 & 4.1 & 6.7 & 2270 & 14 & 0.09 & -8.7 & -7 \\
\hline
\enddata
\tablenotetext{a}{Stellar parameters from \cite{Berlanas2020}.}
\tablenotetext{b}{Stellar parameters from \cite{NebotGomez-Moran2018}.}
\tablenotetext{c}{Stellar parameters from \cite{Rauw2015}.}
\tablenotetext{d}{[O5 V((fc)) + B0:V] + O9V - Spectral type calibration for an O5 V star \cite{Martins2005}.}
\tablenotetext{e}{Stellar parameters from \cite{Marcolino2017}}
\tablenotetext{f}{Stellar parameters from \cite{Carneiro2019}}
\tablenotetext{h}{Stellar parameters from \cite{Berghoefer1996, Martins2005b, Aschenbrenner2023}
\tablenotemark{i}{Stellar parameters from \cite{Wolff2006}}}
\tablenotemark{j}{Stellar parameters from \cite{Pradhan2023}}
\end{deluxetable*}


\section{Conclusions}
\label{conclusions.sec}

\cite{Law2024} presented the first detection of wind emission in high ionization fine-structure lines (\nev, \nevi\, and \ofour) in an O-type star in 10 Lac, but found no trace of the lines in $\mu$~Col. Given the similar spectral types (O9~V and O8.5~V, respectively) this raised the question of whether such wind emission diagnostics would be common among O-type stars or if 10 Lac was a special case.

In this work we have established that high ionization wind signatures in the mid-infrared are not unique to 10~Lac (O9~V) and may be present in O stars across a wider range of spectral types, with confirmed detections of \nev\ wind emission in 4 additional stars (O8.5~V, O8~IV and O5~V), and tentative detections in an another 3 (O8.5~II, O8.5~IV and OC9~Iab). The full sample of 22 stars are comprised of 2 early O stars ($<$O5~V, one confirmed and one tentative detection), 10 late O stars ($>$O8, with a mix of luminosity classes, 3 confirmed and 3 tentative detections) and 10 B stars. We find no evidence for the \nev\ line in any B-type stars. 

Using the line strength of \nev 14.3 \micron\ in the stars with clear detections, we estimate lower limits on mass-loss rates, with the caveats that \added{we assume a 2-level atom when 3 levels are involved in the formation of \nev\,14.32\,$\mu$m, and that the contribution of only collisional excitation populates the upper level.} The mass-loss rate lower limit estimates are in good agreement with theoretical predictions. 
The relative agreement to theoretical predictions is similar to that found in recent literature studies on early O-type stars which use UV and optical diagnostics and include the effects of clumping. 
Using the line widths, we measure terminal wind speeds in good agreement with empirical calibrations (\citealp{Prinja1990, Hawcroft2023}) for 4 stars, while the same calibrations over-predict the wind speeds of 3 stars by at least 700 \kms. 
Using simple geometric models we assess the profile morphology, and find that all but one of the profiles are statistically consistent with the flat-topped profile expected from spherical emission. The line morphology in 
HD~206267 however shows significant deviation from the flat-topped profile, suggestive of enhanced emission in the equatorial plane.

The presence of wind signatures in this sample are particularly notable for two reasons: 

(i) a subset of the sample is obscured by high levels of extinction, prohibiting high quality spectroscopic observations of wind signatures in the UV and optical. These MIR diagnostics therefore allow us to study the winds of previously inaccessible objects. 

(ii) a subset of the detections are for stars in the `weak-wind' regime, offering new insight to the nature of stellar winds for these puzzling objects which have no diagnostic information at UV and optical wavelengths.

Unfortunately the limited composition and non-uniformity of S/N of the current sample of massive stars observed with JWST-MIRI prevents us from making any stronger comment on the presence of additional MIR wind-emission lines, or the likelihood of their presence in OB-type stars in general. We are also unable to make a comparison of wind features in MIR with typical diagnostics at shorter wavelengths due to the limited feasibility of follow-up observations, hampered by the high extinction ($A_{V} > 5$) across the WISCI sample, although there is a viable path towards multi-wavelength modelling of HD~206267 (from MEAD, $A_{V} < 2.5$) if the binary components can be disentangled. 

From the current sample we can conclude that the \nev 14.3 \micron\ line has a diverse range of strengths, widths and morphologies across the sample of now 5 stars with confirmed detections (with an additional 3 tentative), establishing a viable new diagnostic to measure the wind properties of late O-type stars and a new window to investigate wind driving from the MIR. It will soon be possible to study these new MIR diagnostics further with a series of upcoming observations across a variety of late O-type stars that will be obtained as a part of a JWST cycle 4 programme (PID 8683).










\begin{acknowledgments}
We thank Chris Evans, Claus Leitherer, Nimisha Kumari, Frank Backs, Alex de Koter and Lex Kaper for helpful discussions on the nature of MIR emission lines.
\noindent This work is based on observations made with the NASA/ESA/CSA James Webb Space Telescope. The data were obtained from the Mikulski Archive for Space Telescopes at the Space Telescope Science Institute, which is operated by the Association of Universities for Research in Astronomy, Inc., under NASA contract NAS 5-03127 for JWST. 
These observations are associated with programs \#1524, \#2183, \#2459, and \#4497
and can be accessed via doi:10.17909/7qhs-8s19.
MD and SZ acknowledge support from the Research Fellowship Program of the European Space Agency (ESA). DJH gratefully acknowledges support from grant JWST-GO-
07169.003.

\end{acknowledgments}


\bibliography{miri}{}
\bibliographystyle{aasjournalv7}

\begin{appendix}
\renewcommand{\thefigure}{A\arabic{figure}}
\setcounter{figure}{0}

\section{Additional Figures}

\begin{figure*}[!htbp]
\epsscale{1.2}
\plotone{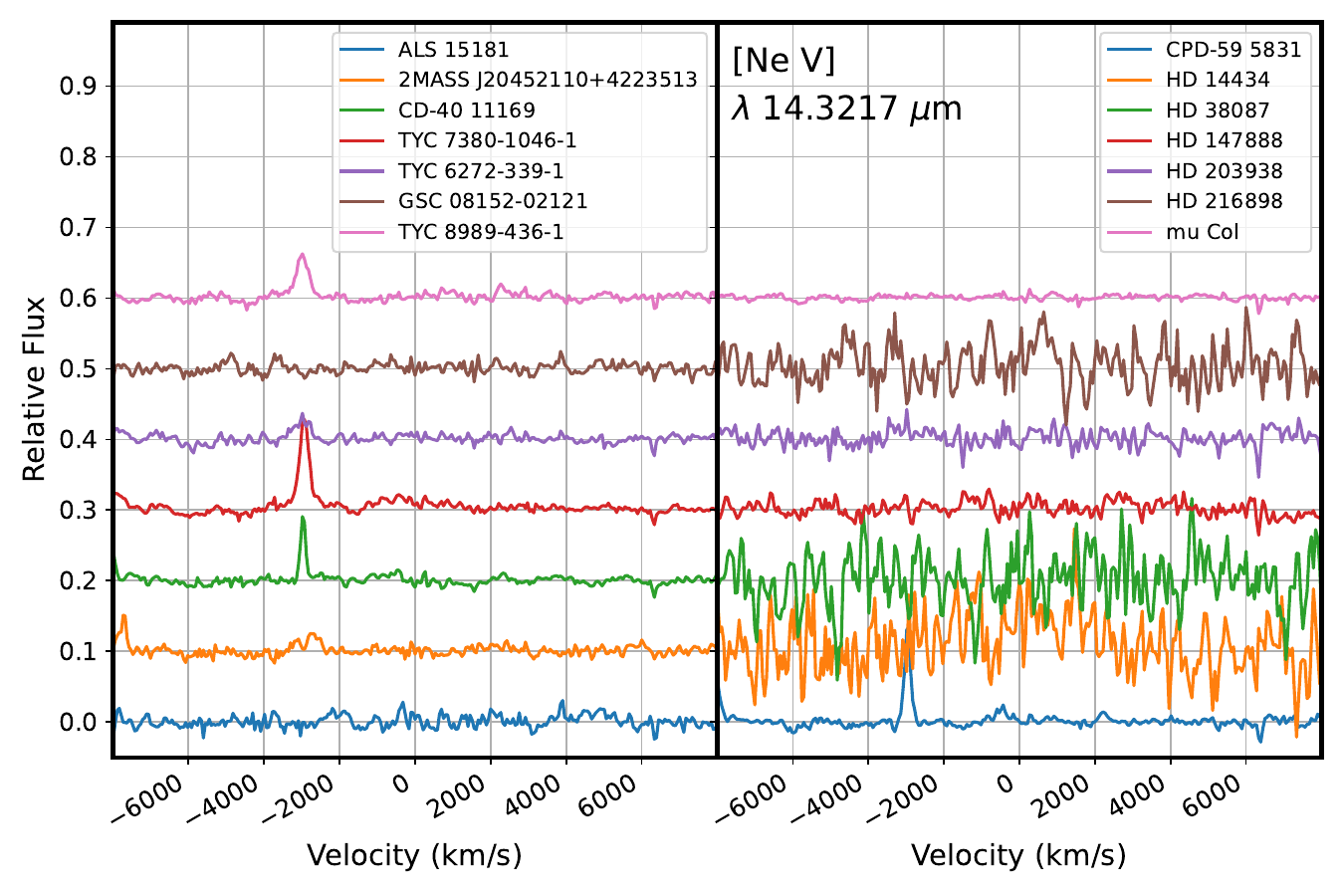}
\caption{As Figure \ref{spectra.fig}, but for the 14 stars in which no statistically significant \nev\ emission was detected.
}
\label{nondet.fig}
\end{figure*}

\begin{figure*}[!htbp]
\epsscale{1.2}
\plotone{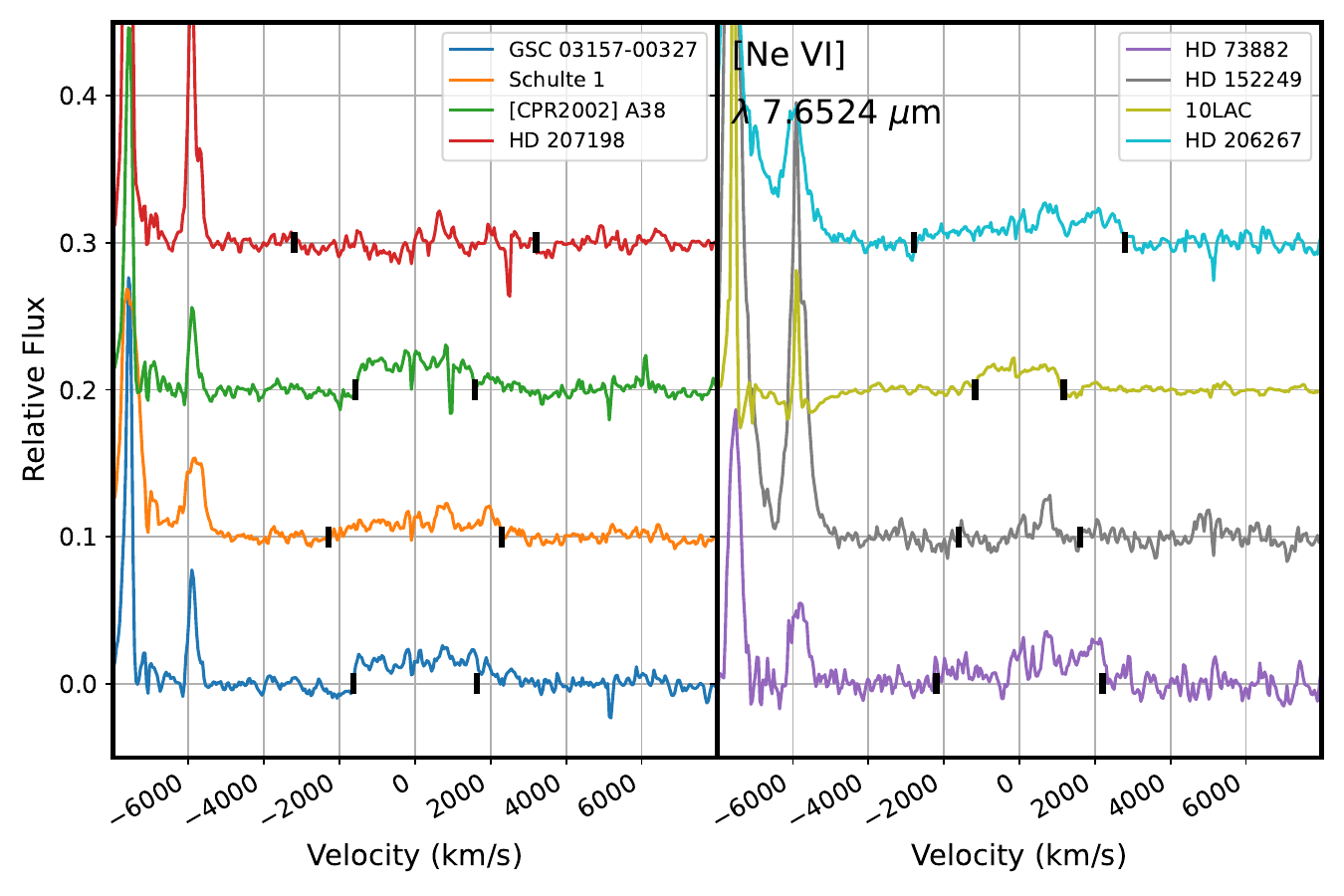}
\caption{As Figure \ref{spectra.fig}, but for \nevi\ emission at $\lambda = 7.6524$ \micron.  The black vertical lines show the maximal wind velocities derived from the \nev\ emission, which is a good match to \nevi\ where such emission can be seen.  The strong emission features at -6000 and -7500 \kms\ are H I 8-6 and 6-5 respectively.
}
\label{nevi.fig}
\end{figure*}

\begin{figure*}[!htbp]
\epsscale{1.2}
\plotone{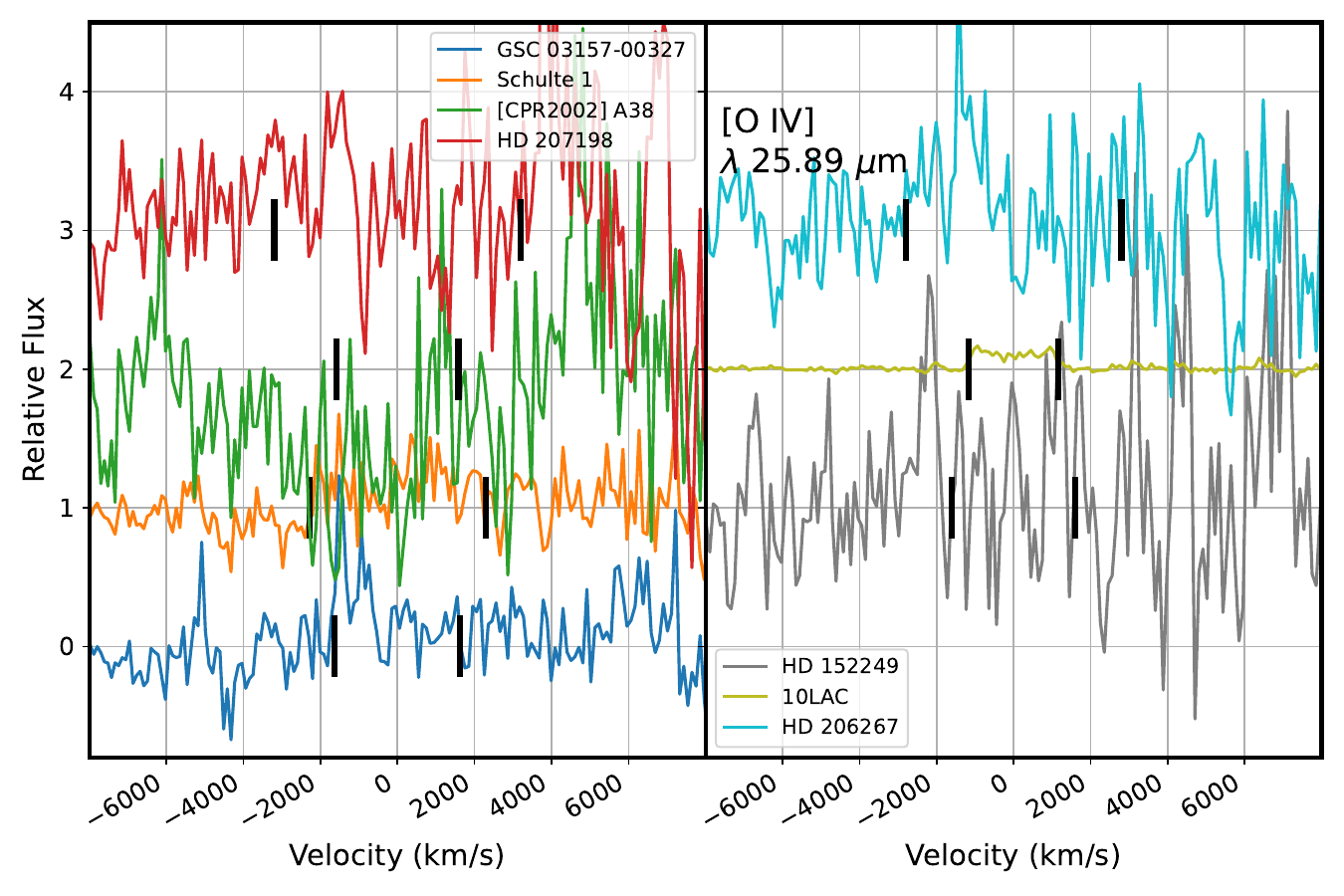}
\caption{As Figure \ref{spectra.fig}, but for [O\four] emission at $\lambda = 25.89$ \micron.  The black vertical lines show the maximal wind velocities derived from the \nev\ emission, which is a good match to [O\four] where such emission can be seen.
}
\label{oiv.fig}
\end{figure*}

Here we show non-detections of NeV in the full sample in Fig. \ref{nondet.fig}. We show the regions around \nevi\ $7.65$ \micron\ and [O\four] $25.89$ \micron\ in Fig. \ref{nevi.fig} and Fig. \ref{oiv.fig}, respectively.

\begin{figure}[!htbp]
\epsscale{1.2}
\plotone{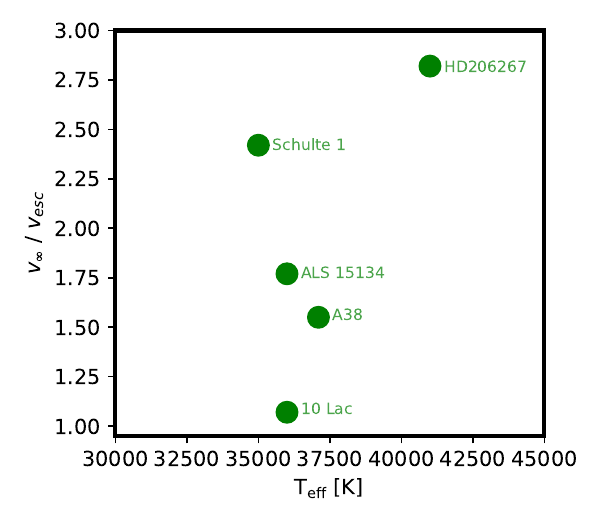}
\plotone{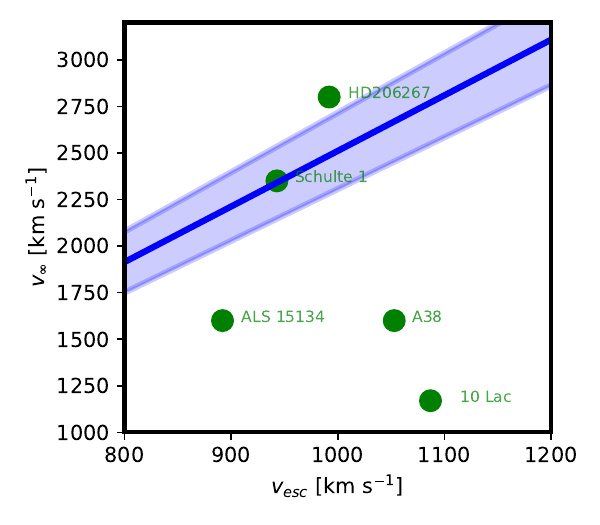}
\caption{Upper panel: Terminal wind speed to escape speed ratio as a function of effective temperature for the 5 stars with strong \nev\ detections. Green points correspond to measurements from the \nev\ 14.3 \micron\ line from this work compared to escape speeds and temperatures taken from the literature. Lower panel: Terminal wind speed as a function of escape speed, points as described for the upper panel. Additionally, the terminal wind speed to escape speed ratio from \cite{Hawcroft2023} is overplotted in blue (with uncertainties included as the shaded region).}
\label{vesc_pred.fig}
\end{figure}

We show trends between terminal wind speed \vinf\ and escape speed $v_{esc}$ in Fig. \ref{vesc_pred.fig}.

\section{Notes on Individual Stars}
\label{notes.sec}

\noindent\textbf{ALS 15134}

ALS 15134 is a late-type dwarf (O8.5~Vz or O8.5~V) with a relatively high extinction ($A_{V} \sim 6$) so there are no existing UV spectra, meaning the wind parameters for this object were previously unconstrained. It is part of the Cygnus OB2 association, and optical spectra are available (e.g. from the Isaac Newton Telescope - INT covering key line profiles from 4000-5000\AA, and including coverage of HeII4686 and H$\alpha$, \citealp{Berlanas2020}). It was classified in the GOSSS catalogue as an O8.5~V$z$ star, suggesting it could be close to the zero age main sequence (ZAMS, \citealp{MaizApellaniz2016}). As discussed further in \cite{Arias2016}, the star crossed the threshold for the $z$ classification due to its relatively strong \ion{He}{2}4686 absorption compared to \ion{He}{1} 4471 (ratio of 1.15 with threshold of 1.1). 
The optical spectrum was then analysed using stellar atmosphere models in \cite{Berlanas2020}, so robust constraints on the fundamental stellar photospheric parameters are available (and listed in Table \ref{tbl:params}). These authors note that ALS 15134 does not have a peculiar proper motion and is not identified as a spectroscopic binary. The rotation rate is moderate ($v$sin$i\sim$ 95\,\kms) and there is no sign of surface helium enrichment, although no metal abundances have been estimated so far. The are no previous spectroscopic studies at longer wavelengths than the optical, but ALS 15134 has been observed in X-rays with XMM-Newton. The X-ray luminosity is relatively low for a typical O-type star at log$L_{X}/L_{bol}$ = -7.1 \citep{NebotGomez-Moran2018}. From the current information available from previous studies this appears to be a normal single O-type star, which would presumably sit within the weak-wind regime based on its late spectral type and luminosity (log$(\frac{L}{L_{\odot}}) \sim$ 5.07).

\medskip
\noindent\textbf{[CPR2002] A38}

A38 is a late-type dwarf (O8~V) with a similar extinction and archival spectra scenario as for ALS 15134. It is also a member of Cygnus OB2 and was previously analyzed by \cite{Berlanas2020}, there is little to add beyond the discussion for ALS 15134 as these are very similar stars, with A38 having a lower luminosity by 0.13 dex, a higher temperature by $\sim$1kK and faster rotation by $\sim$120\kms. One difference is that the X-ray properties for A38 are derived from Chandra observations, both A38 and ALS 15134 are included in \cite{Rauw2015}. These authors note that A38 has an X-ray flux around a factor of 3 lower than ALS 15134.

\medskip
\noindent\textbf{Schulte 1}

Schulte 1 is another late-type dwarf (O8~V) but is notably a spectroscopic binary (SB1) comprised of the O8~V primary and likely a B-type companion (implied by a companion mass of 4-6$M_{\odot}$) \citep{Berlanas2020}, but it does not have reliable astrometry so quantities such as distance and luminosity are not included in that work. The primary mass is around 18 -- 21 $M_{\odot}$ \citep{TriguerosPaez2021}. It is also a short-period (4 days \citealp{Kobulnicky2014}), eccentric eclipsing binary \citep{Prsa2022}, suggested to also show the heartbeat phenomenon (dynamical tidal distortions and tidally induced pulsations). \cite{Wolff2006} estimate a luminosity of log$(\frac{L}{L_{\odot}}) \sim$ 5.35 with a temperature of $\sim 33$\,kK estimated from the spectral type calibrations and bolometric corrections from \cite{Massey2005}. Using the 35\,kK temperature from \cite{Berlanas2020}, along with fluxes and extinction values from \cite{zeegers25} and the bolometric corrections of \cite{Martins2006} gives log$(\frac{L}{L_{\odot}}) \sim$ 5.25. In either case it is unclear whether the observed flux magnitudes are corrected for the contribution of the companion. Therefore it is difficult to tell whether Shulte 1 should be a weak-wind star based on its luminosity, and unfortunately the existing STIS spectra are missing coverage of the key wind features. Schulte 1 has an X-ray flux halfway between ALS 15134 and A38 \citep{Rauw2015}.

\medskip
\noindent\textbf{HD~206267}

HD~206267 is a complex spectroscopic and visual binary system at the center of the Trumpler 37 cluster, detailed in \cite{MaizApellaniz2020}. This object is comprised of a central system Aa (a short period - 3.71 days - SB2 O5~V((fc)) + B0:V binary, designated Aa1 and Aa2) with an Ab component $\sim$100 mas away (an O9~V star), a B component 1.8$\arcsec$ away (which is too faint to contribute significantly to the spectrum), as well as two further away C and D components which are B0 and B0.2 stars.

Given the complicated nature of this stellar system it is difficult to assess the properties of the stars based on previous analyses which have been unable to disentangle the many components, especially as the O9~V Ab star was first classified by \cite{MaizApellaniz2020}. 


The most pertinent issue for this analysis is to identify which component of the system produces the forbidden line emission. Excluding the widest companions, the remaining options are Aa and Ab components. This cannot be resolved with the MIRI observations alone as the 100 mas separation is less than half of the FWHM at the shortest wavelengths. There are however a number of archival UV spectra (from IUE and STIS) showing strong P-Cygni wind profiles in 
{\ion{N}{5} $\lambda$1240} and {\ion{C}{4 $\lambda$1550}
which are formed in the wind of the O5~V (Aa) star. Estimating the wind speed directly and using SEI models through the techniques described in \cite{Hawcroft2023} gives $v_\infty$ = 2815 \kms from CIV 1550 and 2690\,\kms from {\ion{N}{5} $\lambda$1240} which is consistent with 2800 \kms measured from the \nev\ line. 
The STIS spectrum was taken through the $0.1\arcsec \times 0.03\arcsec$ ``Jenkins" slit, which was oriented to exclude any possible contamination from the Ab component. 

This comparison provides some confidence that the O5~V Aa1 star is the primary source of wind emission at all wavelengths in this system. This somewhat complicates the weak-wind driving notion that a sufficiently shock heated wind would be low enough density so as to be inefficient at cooling and remain hot enough into the outer wind that the higher ionization diagnostics appear but the typical diagnostics disappear. However the shock-heated wind simulations of \cite{Lagae2021} only tested an early O-supergiant and late O-dwarf, so the existence of both sets of lines in an early dwarf O star has not been assessed by current simulations. It is also unclear what contribution the O9~V star and/or an interaction between the O5~V and B0~V inner binary has to the wind emission line profiles, a wind-wind collision zone could be another viable mechanism to drive these emission lines. \cite{Pradhan2023} estimate the system is very X-ray bright (log$L_{X}/L_{bol}$ = -6.47). 

\medskip
\noindent\textbf{10 Lac}

10~Lac is a very well studied late-type dwarf (O9~V). A recent compilation of estimates for its effective temperature (T$_{\rm eff}$) and logarithmic gravity (log$g$) was given in Table~6 from \citet{Aschenbrenner2023}, with those authors finding T$_{\rm eff}$\,$=$\,34.55\,$\pm$\,0.3\,kK, and log$g$\,$=$\,4.04\,$\pm$\,0.05.
This is in agreement with the stellar parameters found using a wide range of techniques at different wavelength regimes, as discussed in \citet{Aschenbrenner2023}}. The most extreme range of parameters found through all these studies gives T$_{\rm eff}$\,$=$\,34.6-36.0~kK and log$g$\,$=$\,3.90 -- 4.05 with differences in the approaches and assumptions accounting for discrepancies in best-fit parameters. 10~Lac is moderately nitrogen enriched, radial velocity constant and there is no evidence for line-profile variability. There is has been some uncertainty in the distance to 10~Lac, which is discussed extensively in \citet{Gordon2018} and \citet{Aschenbrenner2023}, with their final estimates now converging to 566\,$\pm$\,59 and 542\,$\pm$\,60\,pc, respectively. It is found to be non-magnetic \citep{David-Uraz2014} and to have an X-ray luminosity typical for O-type stars \citep[logL$_\mathrm{x}$/L$_\mathrm{bol}$\,=\,-7,][]{Berghoefer1996}. 

\medskip
\noindent\textbf{Notes on the \cite{Prinja1990} sample}

From the parameters published in \cite{Prinja1990} it appears as though there are multiple other objects with similar discrepancies in wind speed as seen in A38, ALS 15143 and 10~Lac but measured from the UV. However, upon revisiting archival UV spectra and recent literature, the majority of these stars have received updates to their stellar parameters. For example, for HD~14947 \cite{Prinja1990} quote $v_{\infty}$ as 1885~\kms\ while \cite{Hawcroft2021} measured 2300~\kms. HD~190429A has a very similar correction. HD~14434 on the other hand, has a revised temperature in \cite{Holgado2020} which is 2\,kK lower than used in \cite{Prinja1990}. The same temperature correction is applied to HD~15180 \citep{Crowther2009}. HD~57060 is a contact binary which introduces significant uncertainties on the temperature estimate and may result in a large correction on $v_{\infty}$ if radial velocity variations are not accounted for \citep{AbdulMasih2022}. It is therefore likely that all other outliers in the \cite{Prinja1990} sample (at least above 30\,kK) should be revised to much closer agreement with the general trend. We have corrected these outliers in Fig. \ref{vinf_pred.fig} but they are visible in e.g. \cite{Hawcroft2023}.

\section{Computing the Line-Profiles}\label{AppendixC}

The power from an optically thin emission line received by an external observer in a frequency
interval $d\nu$ centered at frequency $\nu$ is given in a spherical coordinate system by
\begin{align}
L_\nu\, d\nu 
& =  
\int_0^{2\pi} \int_0^\pi \int_0^\infty \eta\, r^2\,\sin\theta\, dr\,d\theta\,d\phi\,d\nu
\label{eqn:linelum}
\end{align}
where $\eta \equiv \eta(r, \theta, \phi, \nu)$ is the volume emissivity 
in {erg\,s$^{-1}$cm$^{-3}$\,\,sr$^{-1}$\,Hz$^{-1}$}. 
For a line transition between upper level $u$ and lower level $l$ with frequency $\nu_{ul}$, 
the emissivity can be separated into position- and frequency-dependent parts:
\begin{eqnarray}
\eta(r, \theta, \phi, \nu) 
& = &
\eta_{ul}(r, \theta, \phi) \times \varphi(\nu - \nu_{ul})  \\
& = &
n_u\,A_{ul} \left(\frac{h \nu_{ul}}{4 \pi}\right) \varphi(\nu - \nu_{ul}) 
\end{eqnarray}
where $n_u(r, \theta, \phi)$ is the number density in the upper level, $A_{ul}$ is the Einstein coefficient
for spontaneous emission, $h$ is Planck's constant, and where the integral of the 
intrinsic profile $\varphi(\nu - \nu_{ul}$) 
over all frequencies is 1.
In the simplified case of a two-level atom in which a collisional excitation from the ground 
state $l=1$ to the excited state $u=2$ at a rate $C_{12}$ is followed either by a spontaneous 
radiative decay or a collisional de-excitation at a rate $C_{21}$, manipulation of the rate 
equation gives
\begin{eqnarray}
\eta_{21}(r, \theta, \phi) 
& = &
n_2\,A_{21}  \left( \frac{h\nu_{21}}{4\pi} \right)  \\
& = &
n_e\,n_1\,C_{12} \left( \frac{1}{1+ n_e/\ncrit} \right) \left( \frac{h\nu_{21}}{ 4\pi} \right) 
\end{eqnarray}
where 
$n_1(r, \theta, \phi)$ is the number density of ions in the ground state, 
$n_e(r, \theta, \phi)$ is the electron number density, 
and the critical density is defined by $\ncrit=A_{21}/C_{21}$.

To explore the effects of aspherical but axisymmetric density distributions on the morphology of 
fine-structure line profiles, we allowed the density of a wind to vary as a function of 
stellar colatitude by an arbitrary, dimensionless function $G(\theta)$ and 
assumed that these changes do not affect the temperature or ionization structure,
which are taken to be constant throughout the wind.
None of these assumptions is likely to be valid, in particular because the ionization balance 
is certainly altered by changes in density.
Nevertheless, these simplifications allow the electron number density to be 
specified as $n_e(r, \theta) = k_e\,G(\theta) / r^2$, where $k_e$ is the mean number of electrons per 
ion in the wind; the number density in the ground state of the ion of interest to be approximated by 
$n_1(r, \theta) \approx n_i(r, \theta) = k_i\,G(\theta) / r^2$, where $n_i$ is the number density 
of the ion and $k_i$ is the fraction of wind material in that ion.  

Finally, we assume that the intrinsic line profile can be replaced by a $\delta$ function:  
$\varphi(\nu - \nu_0) \equiv \delta(\nu - \nu_0)$, which is valid in the Sobolev approximation.  
To account for the shift in line center as seen by an external observer, the profile in a moving medium 
becomes $\delta\left(\nu - \nu_0 \left(1 + v_{LoS}/c \right)\right) \equiv \delta(\Delta\nu_{v_{LoS}})$, 
where $v_{LoS}$ is the velocity of the wind projected along the line-of-sight to the observer.
Since the fine-structure lines are expected to form preferentially in low-density environments at large radii, 
we assume that the velocity is constant, independent of $\theta$ and equal to \vinf, even though in practice 
the velocity will vary with the local mass flux.  
The large emission radius also implies that occultation by the star can be neglected, as already indicated
by the integration limits in Equation~\ref{eqn:linelum}.

With these assumptions, the profile of a fine-structure line is given by:
\begin{equation} 
L_\nu 
 = 
{\int_0^{2\pi}} {\int_0^\pi} {\int_0^\infty} 
\eta_{21}(r, \theta)\, \delta(\Delta\nu_{v_{LoS}})\,
r^2\,\sin{\theta}\,dr\,d\theta\,d\phi .  
\label{eqn:flux}
\end{equation}
The integral over $r$ assumes a standard analytic form with the substitution 
$u= \sqrt{\ncrit/ G(\theta)\,k_e}\,r$:
\begin{eqnarray}
\int_0^\infty \eta_{21}(r, \theta)\, r^2 dr 
& = & 
C\,G^{\frac{3}{2}}(\theta) 
\int_0^\infty 
\frac{1}{u^2 + 1}\,du  \\
& = &
C\,G^{\frac{3}{2}}(\theta)\,\left(\frac{\pi}{2}\right) 
\end{eqnarray}
where the constant
\begin{equation}
C = 
\left(k_e n_{crit}\right)^{\frac{1}{2}}\,k_i\,
C_{12}\,\left({h\nu_{21}/4\pi}\right) .
\end{equation}
Consequently, Equation~\ref{eqn:flux} simplifies to
\begin{equation}
 L_\nu  = C\,\left(\frac{\pi}{2}\right) \int_0^{2\pi} \int_0^\pi G^{\frac{3}{2}}(\theta) \,
\delta(\Delta\nu_{v_{LoS}})
\sin{\theta}\,d\theta\,d\phi.
\label{eqn:flux2}
\end{equation}
Note that the integrand in Equation~\ref{eqn:flux2} is 0 except when $\Delta\nu_{v_{LoS}} = 0$; 
i.e., when $\nu = \nu_{21} \left(1 + v_{LoS}/c \right)$, which simplifies the integration.
Also note that $L_\nu$ can be compared directly to the observed flux $F_\nu$, since
conservation of energy requires
\begin{equation}
L_\nu = 4\pi D^2 F_\nu,
\end{equation}
where $D$ is the distance to the star and the units of flux are 
{erg\,s$^{-1}$\,cm$^{-2}$\,Hz$^{-1}$}.

\begin{figure}[!ht]
\epsscale{1.0}
\plotone{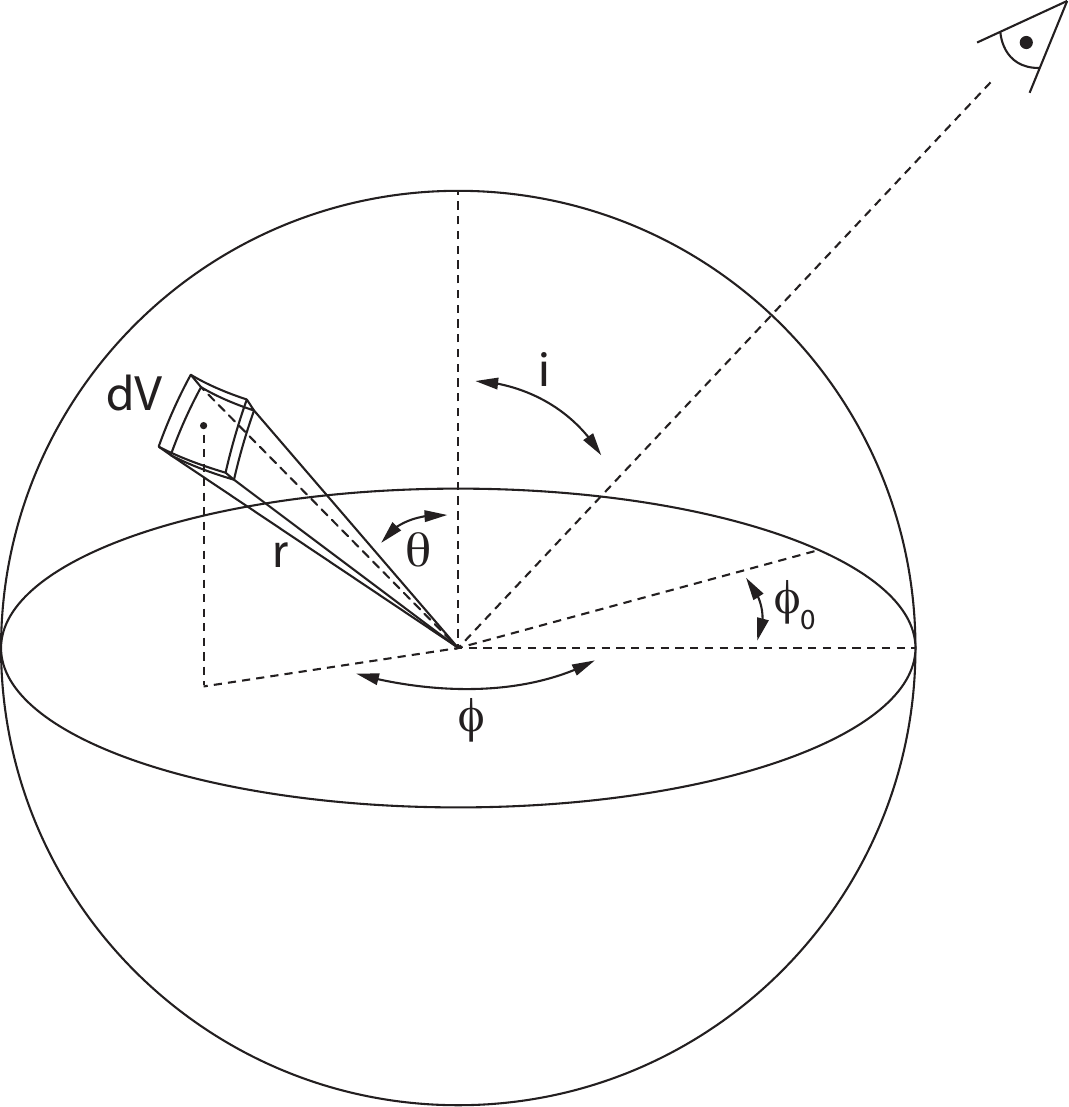}
\caption{Schematic illustration of the spherical coordinate system adopted for the wind model.  
         The rotational axis of the star is aligned with the direction $\theta = 0$ and the wind 
         is viewed from an angle $(\theta_0, \phi_0) = (i, 0)$.
        }
\label{diagram.fig}
\end{figure}
Figure~\ref{diagram.fig} illustrates the geometry needed to determine $v_{LoS}$.
For an axisymmetric configuration, the observer can be placed in the x-z plane without 
loss of generality.
By aligning the rotational axis of the star with the $\theta = 0$ direction, the observer's
line of sight in Cartesian $(\hat i, \hat j, \hat k)$ coordinates is 
\[
\hat n_{obs} = \sin{i}\,\,\hat i  + \cos{i}\,\,\hat k
\]
and the direction of a patch of the radial outflow is given by 
\[
\hat{r}  = \sin{\theta}\,\cos{\phi}\,\,\hat i  +  \sin{\theta}\,\sin{\phi}\,\,\hat j + \cos{\theta}\,\,\hat k
\]
Consequently
\begin{eqnarray}
v_{LoS} & = & v_\infty\,(\hat n_{obs}.\hat{r})  \\
        & = & v_\infty\,(\sin{i}\,\sin{\theta}\,\cos{\phi}  + \cos{i}\,\cos{\theta})
\label{eqn:vlos}
\end{eqnarray}

Flux profiles were constructed by numerically integrating Equation~\ref{eqn:flux2} on a discrete mesh 
of $(\theta, \phi)$ points spaced at intervals of $(\Delta\theta, \Delta\phi) = (0.2^\circ, 0.2^\circ)$.  
For a specified inclination angle $i$, each patch at $(\theta_j, \phi_k)$ 
contributes its surface area weighted by $G(\theta_j)$ to the flux profile at the value of $v_{LoS}$ 
given by Equation~\ref{eqn:vlos}.
The fluxes were accumulated in bins with widths of 10~\kms to produce the profiles illustrated in 
Figure~\ref{geometry.fig}.

As a final caveat, we note that this approach assumes a 2-level atom and that collisions dominate the 
population of the upper level.
The 2-level atom approach is valid for {\nevi}\,7.65\,$\mu$m and {\ofour}\,24.89\,$\mu$m lines, 
but 3 levels are involved in the formation of the {\nev}\,14.32\,$\mu$m and 24.89\,$\mu$m features.
Pumping by UV continuum radiation may also play a role in populating the upper level of these transitions.



\end{appendix}

\end{document}